\documentclass[conference, a4paper]{IEEEtran}
\usepackage[left=1.62cm,right=1.62cm,top=1.9cm]{geometry}

\usepackage[utf8]{inputenc} 
\usepackage[T1]{fontenc}
\usepackage{url}              
\usepackage{cite}             

\usepackage[cmex10]{amsmath}  
\usepackage{mleftright}       
\mleftright                   

\usepackage{graphicx}         
\usepackage{booktabs}         

\usepackage{float}
\usepackage{booktabs}
\usepackage{units}
\usepackage{amsmath}
\usepackage{amsthm}
\usepackage{graphicx}
\usepackage{color}
\makeatletter

\providecommand{\tabularnewline}{\\}
\floatstyle{ruled}
\newfloat{algorithm}{tbp}{loa}
\providecommand{\algorithmname}{Algorithm}
\floatname{algorithm}{\protect\algorithmname}

\theoremstyle{plain}

\theoremstyle{plain}

\makeatother
\providecommand{\corollaryname}{Corollary}
\providecommand{\theoremname}{Theorem}
\usepackage{geometry}
\begin{document}

\title{Fast Blind Recovery of Linear Block Codes over Noisy Channels}
\author{%
  \IEEEauthorblockN{Peng Wang$^{\star}$, Yong Liang Guan$^{\ddagger}$, Lipo Wang$^{\ddagger}$, and Peng Cheng$^{\star\dagger}$}
  \IEEEauthorblockA{%
    $^{\star}$ School of Electrical and Information Engineering, The University of Sydney\\
    $^{\ddagger}$ School of Electrical and Electronic Engineering, Nanyang Technological University, Singapore\\
    $^{\dagger}$ Department of Computer Science and Information Technology, La Trobe University, Australia\\
    Email: wpwd1986@hotmail.com, \{EYLGuan, ELPWang\}@ntu.edu.sg, p.cheng@latrobe.edu.au}
}



\maketitle 
\begin{abstract}
This paper addresses the blind recovery of the parity check matrix of an $(n,k)$ linear block code over noisy channels by proposing a fast recovery scheme consisting of 3 parts. Firstly, this scheme performs initial error position detection among the received codewords and selects the desirable codewords. Then, this scheme conducts Gaussian elimination (GE) on a $k$-by-$k$ full-rank matrix and uses a threshold and the reliability associated to verify the recovered dual words, aiming to improve the reliability of recovery. Finally, it performs decoding on the received codewords with partially recovered dual words. These three parts can be combined into different schemes for different noise level scenarios. The GEV that combines Gaussian elimination and verification has a significantly lower recovery failure probability and a much lower computational complexity than an existing Canteaut-Chabaud-based algorithm, which relies on GE on $n$-by-$n$ full-rank matrices. The decoding-aided recovery (DAR) and error-detection-\&-codeword-selection-\&-decoding-aided recovery (EDCSDAR) schemes can improve the code recovery performance over GEV for high noise level scenarios, and their computational complexities remain much lower than the Canteaut-Chabaud-based algorithm. \end{abstract}


\thispagestyle{empty}

\section{\label{sec:Introduction}Introduction}

In many modern advanced telecommunication systems, including adaptive modulation and coding (AMC) \cite{Han06A,Ishii01A,Goldsmith98A} and cognitive radio systems \cite{Marazin09D}, blind recognition of the modulation, coding, or scrambling parameters of an unknown incoming communication signal waveform, and the subsequent use of the recovered parameters to detect and decode the incoming signal, are indispensable \cite{Moosavi14F,Yardi16B,Swam17C}. More specifically, control channels are needed to transmit the AMC parameters to implement the AMC in mobile network systems. To preserve the spectrum usage, a technique called blind decoding was proposed to blindly estimate the AMC parameters from the received noisy data from the data channel without the usage of control channels. This paper focuses on the blind identification of the parity check matrix of an $(n,k)$ linear block code without a candidate set in a noisy channel, where $k$ and $n$ are the number of message bits and coded binary bits,  respectively.

To blindly identify the parity check matrix of a $(n,k)$ linear block code in a noisy channel, a sufficient number, that is, $n-k$, of dual words that are orthogonal to the codewords need to be recognized. In \cite{Cluzeau09R,Cluzeau06B,Cluzeau09R2}, techniques for linear block code recovery based on exhaustive search were proposed. These brute force search (BFS) schemes typically used a threshold to select the correct dual words out of all $2^{n}$ possible dual words, leading to computational complexity increases exponentially with the code length $n$. In \cite{Valembois01D}, Valembois proposed a method to find the correct dual words based on the Canteaut-Chabaud information set decoding algorithm \cite{Canteaut98A}, and recently Cluzeau \cite{Cluzeau09R,Cluzeau06B} and Cote \cite{Cote09R} used this technique for the reconstruction of different types of codes. However, these schemes need to perform Gaussian elimination (GE) on an $n$-by-$n$ full-rank matrix and a small-scale exhaustive search of all combinations of $2p$ columns out of $n$ columns, where $p$ is usually either one or two. In \cite{Yu17L}, an algorithm based on the Canteaut-Chabaud algorithm \cite{Canteaut98A} to find the low-weight dual words from the dual space of the received code-vector space to recover LDPC codes. In \cite{Carrier19I}, an algorithm based on the partial Gaussian elimination algorithm to find the sparse part in the noisy matrix to recover LDPC codes. However, their methods focused on the recovery of LDPC codes which has the special property of a sparse parity-check matrix. This paper focuses on the recovery of all linear block codes. 

This paper first proposes a novel scheme to recover the dual words of a linear block code with a significantly reduced computational complexity.  This scheme performs initial error position detection among the received codewords and selects the codewords with the least detected errors for the coding recovery process. The proposed scheme conducts GE on a $k$-by-$k$ full-rank matrix to recover the dual word candidates. Then, the reliability associated with each dual word candidate is used in a verification process to improve the reliability of recovery when the number of received codewords is limited. After the verification, $n-k$ linearly independent dual words with the highest reliability will be selected as the recovered dual words. The early-recovered parity check rows to clean the noisy received codewords to enhance the recovery of the remaining parity check rows. These three parts can be combined into different schemes for different noise level scenarios. The GEV that combines Gaussian elimination and verification has a significantly lower recovery failure probability and a much lower computational complexity than an existing Canteaut-Chabaud-based algorithm, which relies on GE on $n$-by-$n$ full-rank matrices. The decoding-aided recovery (DAR) and error-detection-\&-codeword-selection-\&-decoding-aided recovery (EDCSDAR) schemes can improve the code recovery performance over GEV for high noise level scenarios, and their computational complexities remain much lower than the Canteaut-Chabaud-based algorithm.



\section{\label{sec:Network-Model-and}System Model and Problem Formulation}

In this section, we introduce the basic system model for the transceivers. We consider a general communication system with the additive white Gaussian noise (AWGN) channel model.

This paper conducts all algebraic operations and the associated analysis in binary field $GF(2)$. The original information bits are grouped into blocks at the transmitter, each of which consists of $k$ consecutive bits. The $v$-th information bits block $\mathbf{s}_{v}$ is passed to an unknown linear block encoder $\mathcal{C}$ to generate a corresponding codeword $\mathbf{c}_{v}$. Specifically, we have
\begin{align}
\mathbf{c}_{v} & =\mathbf{s}_{v}\mathbf{G},\label{eq:1}
\end{align}
where $\mathbf{c}_{v}=(c_{v}^{1},...,c_{v}^{n})\in GF(2)^{n}$ is the $1$-by-$n$ codeword vector, $\mathbf{s}_{v}=(s_{v}^{1},...,s_{v}^{k})\in GF(2)^{k}$ is the $1$-by-$k$ information vector, $\mathbf{G}=(\mathbf{I}_{k},\mathbf{p}_{1},\cdots,\mathbf{p}_{n-k})$ is the $k$-by-$n$ generator matrix of the linear block encoder $\mathcal{C}$ which is unknown to the receiver and needs to be recovered. $\mathbf{I}_{k}$ is the identity matrix with size $k$ and $\mathbf{p}_{i},i=1,...,n-k$ are $k$-by-$1$ columns. The corresponding code rate is $R=k/n$. The codeword $\mathbf{c}_{v}$ is assumed to be modulated by a binary phase-shift keying (BPSK) modulator and the corresponding block of modulated symbols is denoted by $\mathbf{b}_{v}=(b_{v}^{1},...,b_{v}^{n})\in\mathcal{B}^{n}$, where $\mathcal{B}=\left\{ -1,+1\right\} $. The soft values of the received bits are also collected in blocks $\mathbf{y}_{v}=(y_{v}^{1},...,y_{v}^{n})$ and can be expressed as 
\begin{eqnarray}
y_{v}^{j} & = & b_{v}^{j}+w_{v}^{j},j=1,2,...,n,\label{eq:2-1}
\end{eqnarray}
where under the AWGN channel model, the values of the independent
noise $w$ follow the centered-Gaussian probability distribution function
\begin{eqnarray}
p(w) & = & \frac{1}{\sqrt{2\pi\sigma^{2}}}\exp\left[\frac{-w^{2}}{2\sigma^{2}}\right],\label{eq:3-1}
\end{eqnarray}
with $\sigma^{2}=\nicefrac{N_{0}}{2}$, where $N_{0}$ is the noise
spectral density.

Let $\oplus$ denote the addition over binary field $GF(2)$ (or exclusive-OR (XOR) operation). After demodulation, the hard decisions of the bits
in blocks $\mathbf{r}_{v}=(r_{v}^{1},...,r_{v}^{n})$
can be represented as 
\begin{eqnarray}
\mathbf{r}_{v} & = & \mathbf{c}_{v}\oplus\mathbf{e}_{v},\label{eq:3-2}
\end{eqnarray}
where $\mathbf{e}_{v}=(e_{v}^{1},...,e_{v}^{n})\in GF(2)^{n}$
is the random variables with the crossover probability $P_{e}$ defined by 
\begin{eqnarray}
P_{e} & = & \frac{1}{2}\mbox{erfc}\left(\frac{1}{\sqrt{2\sigma^{2}}}\right),\label{eq:4-1}
\end{eqnarray}
where $\mbox{erfc}()$ is the complementary error function.

In this paper, we assume a time synchronization has been achieved and the
values of $n$ and $k$ have been found, e.g., using techniques in
\cite{Sharma16B}\footnote{By using the method proposed in \cite{Sharma16B}, we can obtain the correct values of $n$ and $k$ when the SNR is above a certain value. For the received noisy bitstream, we can use our proposed error detection and codeword selection method to eliminate the noise to reduce the SNR.}. Then the remaining problem of blind recovery of
the linear block code is to find a parity check matrix of rank $n-k$
from the noisy observations. Let $\mathcal{C}^{\bot}$ denotes the
dual of a code $\mathcal{C}$, i.e., 
\begin{equation}
\mathcal{C}^{\bot}=\{\mathbf{h}|\forall\mathbf{c}\in\mathcal{C},\mathbf{c}\cdot\mathbf{h}=0\},\label{innerhc-1}
\end{equation}
where $\mathbf{c}\cdot\mathbf{h}$ denotes the inner product of $\mathbf{h}$
and $\mathbf{c}$ in $GF(2)$. Recovery of the $(n,k)$ linear block
code can be achieved by finding the $n-k$ dual words $\mathbf{h}$
in $\mathcal{C}^{\bot}$ that are linearly independent.



\section{\label{sec:EDCSDAR}A Low Complexity Algorithm with Error Detection and codeword selection and Decoding Aided-Recovery and Gaussian Elimination and Verification}

In this section, a new scheme, referred to as Error Detection and codeword selection and Decoding Aided-Recovery (EDCSDAR) scheme, which is much simpler than the existing algorithms described in Appendix A, is proposed for linear block code recovery. The proposed EDCSDAR algorithm includes three key steps: 1) error position detection and codeword selection;  2) Gaussian elimination (GE) and verification of the recovered dual words; 3) noise-reducing decoding.

\subsection{\label{sec:EDCSDARsub1}Error Detection and codeword selection}

Recall that we consider the AWGN channel model. The codewords are assumed to be modulated by a BPSK modulator. So each received bit has a Gaussian noise in it. The received
noisy soft value stream is firstly divided into $M$ blocks of length $n$ and then arranged in an $M$-by-$n$ matrix $\mathbf{Y}$. After demodulation, the hard decisions of matrix $\mathbf{Y}$ is an $M$-by-$n$ binary matrix $\mathbf{X}$.

It is observed that when a received bit's soft value is near 0, it is more likely that there is an error in the hard demodulated bit, i.e., if the absolute value of a received bit is smaller than others, it is more likely that an error occurs during the demodulation.

With this observation into consideration, we can estimate the error positions in the hard received codeword matrix with a certain probability, select the codewords in which the positions of the detected errors concentrate in a certain number of rows corresponding to the bits of a received codeword or with least number of errors. In this way, the correct dual words that are not affected by the detected errors should be obtained with the code recovery scheme.

Recall that the parity check matrix $\mathbf{H}=\left(\mathbf{P},\mathbf{I}_{(n-k)}\right)$ has a systematic formation, each dual word has more non-zero elements in the first $k$ positions and only $1$ non-zero element in the rest $n-k$ positions. Hence, the errors should be more concentrated in the $n-k$ positions where the non-zero elements are sparse. 

\begin{enumerate}

\item Divide received soft codeword matrix $\mathbf{Y}$ into two sub-matrices $\mathbf{Y}_{1}$ with size $M$-by-$k$ and $\mathbf{Y}_{2}$ with size $M$-by-$(n-k)$.
\item Conduct error detection on $\mathbf{Y}_{1}$ with $\left|\mathbf{Y}_{1}\right|\leq t_{1}$ and $\mathbf{Y}_{2}$ with $\left|\mathbf{Y}_{2}\right|\leq t_{2}$  to obtain the estimated error sub-matrices $\mathbf{E}_{1}$ and $\mathbf{E}_{2}$. 
\item Concatenate two estimated error matrices to form the $M$-by-$n$ complete estimated error matrix $\mathbf{E}=(\mathbf{E}_{1},\mathbf{E}_{2})$.
\item Find the indexes of the columns of $\mathbf{E}$ that have all zeros. The corresponding columns in the received codeword matrix $\mathbf{X}$ have no estimated errors in them, which are selected to form $M_s$-by-$n$ $\mathbf{X}_s$ used to recover some dual words.  

\end{enumerate}

\subsection{\label{sec:Improved-Algorithm-with}
Gaussian Elimination and Verification}

In this subsection, a new scheme, referred to as  Gaussian elimination with verification (GEV) is proposed for linear block code recovery. The proposed GEV algorithm includes two key steps. The first step is to recover the dual word candidates by performing GE on a $k$-by-$k$ full-rank matrix. The second step is to verify the recovered dual word candidates by using the reliability associated with each dual word candidate to improve recovery reliability. The proposed scheme finally selects $n-k$ linearly independent dual words with the highest reliabilities as the recovered dual words. Details of the proposed algorithm will be described in the following 

\begin{enumerate}
\item Divide the matrix $\mathbf{X}_s$ into two sub-matrices $\mathbf{X}_{s1}$ with size $M_s$-by-$k$ and $\mathbf{X}_{s2}$ with size $M_s$-by-$(n-k)$, so that $\mathbf{X}_s=\left(\mathbf{X}_{s1},\mathbf{X}_{s2}\right)$.
\item Perform GE on the $k$-by-$M_s$ matrix $\mathbf{X}_{s1}^{T}$ with Gauss\textendash Jordan elimination through pivoting (GJETP) algorithm \cite{Golub96M} to find the column indexes of the columns with pivots to obtain the $k$-by-$k$ matrix $\left(\mathbf{R}_{1}\right)^{T}$ and obtain the corresponding transition matrix $\mathbf{D}_{t1}$ as the inverse matrix of the full-rank matrix $\left(\mathbf{R}_{1}\right)^{T}$. With the column indexes of the columns with pivots, we can select the corresponding columns in $\mathbf{X}_{s2}^{T}$ to obtain $\left(\mathbf{R}_{2}\right)^{T}$ with size $(n-k)$-by-$k$ 
\item Obtain $\widehat{\mathbf{p}}_{j}$ which is defined as the recovered $\mathbf{p}_{j}$ with
\begin{eqnarray}
\mathbf{\hat{p}}_{j} & = & \left(\mathbf{R}_{1}\right)^{-1}\mathbf{r}_{2,j}.\label{eq:4.1.3}
\end{eqnarray}
where $\mathbf{r}_{2,j}$ is the $j^{th}$ column in the matrix $\mathbf{R}_{2}$
\item Obtain the recovered dual words, i.e., the rows of the obtained parity check matrix $\hat{\mathbf{H}}=\left(\hat{\mathbf{P}},\mathbf{I}_{(n-k)}\right)$, where $\hat{\mathbf{P}}=\left(\hat{\mathbf{p}}_{1},\cdots,\hat{\mathbf{p}}_{n-k}\right)$ is the recovered $\mathbf{P}$ matrix. 
\item For each obtained dual word \textbf{$\mathbf{h}_{t}$}, calculate $\mathbf{w}_{\mathbf{h}_{t}}=\mathbf{h}_{t}\mathbf{X}^{T}$, where $\mathbf{w}_{\mathbf{h}_{t}}$ has length $M$ and weight $d_{\mathbf{h}_{t}}$. 
\item If $d_{\mathbf{h}_{t}}\leq T_{\mathbf{h}}$, where $T_{\mathbf{h}}$ can be calculated with equation (\ref{eq:valueT}), we calculate the reliability of each dual word, which denotes the probability that a tested $n$-tuple is a true dual word \cite{Valembois01D} by
\begin{eqnarray}
\!\!\!\!\!\!\!\!\!\!\!\!& &\!\!\!\!\!\! p_{\mathbf{h}_{t}}\nonumber \\
\!\!\!\!\!\!\!\!\!&=&\!\!\!\!\!\!\frac{\left(1\!+\!(1\!-\!2P_{e})^{\left|\mathbf{h}_{t}\right|}\right)^{M-d_{\mathbf{h}_{t}}}\left(1\!-\!(1\!-\!2P_{e})^{\left|\mathbf{h}_{t}\right|}\right)^{d_{\mathbf{h}_{t}}}}{2^{M}}.\label{probability-3}
\end{eqnarray}
\item Use $\mathbf{h}_{t}$ and $p_{\mathbf{h}_{t}}$ to update the dual word table ($\mathbf{DWT}$) which stores maximally ($n-k$) candidates of dual words. These dual word candidates must be linearly independent. 
\item After all obtained dual word candidates are tested, take the $n-k$ dual word candidates in $\mathbf{DWT}$ as the recovered dual words and use them to form the parity check matrix. 
\end{enumerate}

After the verification, to make sure that the newly obtained dual word candidates are not correlated with the previously obtained dual word candidates in table $\mathbf{DWT}$, we need to use GE again. For this purpose, we propose another efficient GE algorithm, which allows us to use the knowledge of previously obtained linearly independent dual word candidates through pivoting. The details of this algorithm are presented in Appendix B.

\subsection{\label{sec:Further-Performance-Improvement}Noise-Reducing Decoding}

After recovering $n-k$ dual words, the parity check matrix can be reconstructed and used to perform error correction. For some error-correcting codes, e.g., LDPC codes \cite{Gallager63L,Luby98I}, error correction can be done without recovering all the $n-k$ dual words. When some of the dual words are missing, the LDPC code still has some error correction capability and the BER decreases with a decreasing number of missing dual words.

Based on the above analysis, we propose a decoding-aided recovery (DAR) scheme that uses early-recovered parity check rows to clean up the noisy received codewords to enhance the recovery of the remaining parity check rows. This algorithm includes the following steps: 
\begin{enumerate}
\item Conduct code recovery procedure with the proposed GEV on the $M$-by-$n$ matrix $\mathbf{X}$ to obtain $i$ linearly independent dual words to form the matrix $\mathbf{H}^{'}$. 
\item If $1\leq i \leq n-k-1$, conduct decoding for each individual received noisy soft codeword $\textbf{y}_i$ with the obtained dual word matrix $\mathbf{H}^{'}$ with adaptive belief-propagation decoder to obtain the decoded bit $\textbf{y}_i^{'}$ to form the $M$-by-$n$ decoded codeword matrix $\mathbf{X}^{'}$.
\item Conduct code recovery procedure with the proposed GEV on the $M$-by-$n$ decoded codeword matrix $\mathbf{X}^{'}$ to obtain $i^{'}$ linearly independent dual words to form the $i^{'}$-by-$n$ dual word matrix $\mathbf{H}^{'}$. 
\item Proceed to Step 2 until $N_{iter}$ iterations have reached.
\end{enumerate}
The proposed DAR scheme requires at least one recovered dual word to commence the decoding process. For highly noisy channels, this scheme will not perform well.

\subsection{Different Schemes for Different Noise Scenarios}
The GEV that combines Gaussian elimination and verification has a significant performance improvement compared to that achieved by an existing algorithm in terms of recovery accuracy, and the computational complexity is also much lower than the existing ones. Simulation results show that using the proposed GEV algorithm can improve the code recovery performance when a small number of GE is used and the noise level is relatively small. For higher noise levels, one can only recover some of the correct dual words, but not all of $n-k$ correct dual words,  the decoding-aided recovery (DAR) combining GEV and noise-reducing decoding can improve the code recovery performance over GEV. For even more more severe noise levels, error-detection-\&-codeword-selection-\&-decoding-aided recovery (EDCSDAR) schemes combining error-detection-\&-codeword-selection, GEV, and noise-reducing decoding can improve the code recovery performance over GEV and DAR. After more than one dual words are recovered with the aid of error detection and codeword selection, one can use the DAR scheme to decode all the selected codewords and obtain the decoded log-likelihood ratio (LLR) values of all the received codewords, which are used to estimate error positions with LLR thresholds $\nicefrac{4\times t_{1}}{N_{0}}$ and $\nicefrac{4\times t_{2}}{N_{0}}$ in the next iteration, where $N_{0}$ is the noise
spectral density. Then, the estimated error matrix and the selected decoded codewords are updated accordingly to recover more dual words. The DAR scheme's and EDCSDAR scheme's computational complexities remain much lower than the Canteaut-Chabaud-based algorithm.

\subsection{Analysis of the Computational Complexity}

Note that the computational complexity of the proposed GEV algorithm is mainly related to the GE process on a $k$-by-$k$ matrix, and there are $N_{GE}$ times of GE. Hence, the computational complexity of the proposed GEV algorithm is $O(N_{GE}\times M \times k^{2})$.

The computational complexity of the DAR algorithm is mainly related to several iterations of dual word recovery using the proposed GEV algorithm and channel decoding with the recovered dual words. It can be observed that the computational complexity of channel decoding with the part of dual words can be neglected as compared with the computational complexity of the dual word recovery process. Note that there are $N_{iter}$ number of iterations used. Therefore, the computational complexity of the DAR algorithm is $O(N_{iter}\times N_{GE}\cdot M\cdot k^{2})$. The main contributing factor to the computational complexity of EDCSDAR is several iterations of error position detection and channel decoding with the recovered dual words on all the received codewords. Therefore, the computational complexity of the EDCSDAR algorithm is $O(N_{iter}\times(N_{GE}\cdot M\cdot k^{2}+Mn))$.

The computational complexities of the existing algorithms, including the Brute Force Search and the Canteaut-Chabaud algorithms, are compared with that of the proposed GEV algorithm and DAR algorithm, and the results are shown in Table I. 
\begin{table}[t]
\caption{The computational complexity of the code recovery algorithms. }

\centering{}%
\begin{tabular}{cc}
\toprule 
Algorithm  & Complexity\tabularnewline
\midrule 
\multicolumn{1}{c}{Brute Force Search} & $O(M 2^{n})$\tabularnewline
\multicolumn{1}{c}{Canteaut-Chabaud} & $O(N_{GE} (M n^{2}+\left(_{\:\:p}^{n/2}\right)^{2}))$\tabularnewline
\multicolumn{1}{c}{Yu's scheme} & $O\left(N_{c2} \left[M n^{2}+\left(N_{c1}\right.\right.\right.$ \tabularnewline
\multicolumn{1}{c}{} & $\left.\left.\left.\left(n (n-k)^2+\left(_{\:\:p}^{(n-k)/2}\right)^{2}\right)\right)\right]\right)$\tabularnewline
Proposed GEV  & $O(N_{GE}\cdot M\cdot k^{2})$\tabularnewline
\multicolumn{1}{c}{Proposed DAR} & $O(N_{iter}\cdot N_{GE}\cdot M\cdot k^{2})$\tabularnewline
\multicolumn{1}{c}{Proposed EDCSDAR} & $O(N_{iter}(N_{GE}Mk^{2}+Mn))$\tabularnewline
\bottomrule
\end{tabular} 
\end{table}
From Table I, it is shown that the proposed algorithms reduce the computational complexity significantly as compared with the existing ones.


\section{\label{sec:Simulation-Results}Simulation Results }

To demonstrate the performance of the proposed GEV algorithm, it is used to blindly recover a linear block code, i.e., cyclic (100, 50) code. The recovery process is considered to be successful if $n-k$ linearly independent dual words are found. If the number of linearly independent dual words found is less than $n-k$, or there are falsely detected dual words in the $n-k$ dual words, or the recovery process cannot be started because a $k$-by-$k$ or an $n$-by-$n$ full-rank matrix cannot be found, the recovery is considered to be a failure. The channel is an AWGN channel, and we assume that the noise variance $\sigma^{2}$ is known. If $\sigma^{2}$ is unknown, we can use some existing techniques, e.g., the expectation-maximization (EM) estimator proposed in \cite{Xia14N} to estimate it. The modulation scheme is BPSK scheme with the symbol power being 1, the channel error probability $P_{e}$ is given by \eqref{eq:4-1}.

\begin{figure}[!t]
	\setlength{\abovecaptionskip}{-0.05cm}
\centering \includegraphics[width=8cm]{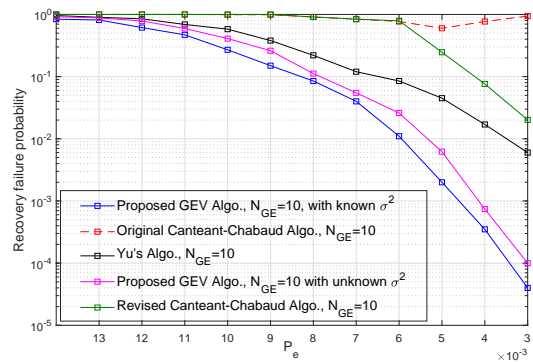} \caption{Recovery failure probabilities of the proposed GEV algorithm, Canteaut-Chabaud
algorithm and Yu's algorithm for the Cyclic (100, 50) code at different values of $P_{e}$. }
\label{errorBSC} 
\end{figure}

The recovery failure probabilities of the proposed GEV algorithm, original Canteaut-Chabaud algorithm, revised Canteaut-Chabaud algorithm, and Yu's algorithm for the cyclic (100,50) code are shown in Fig.~\ref{errorBSC}. It can be observed that the original Canteaut-Chabaud algorithm gets a very high recovery failure probability at a low channel error probability ($P_{e}\leq5.0\times10^{-3}$). This is because the Canteaut-Chabaud algorithm relies on conducting GE on an $n$-by-$n$ full-rank matrix. For a $(n,k)$ linear block codes, all the codeword is among a vector space $span(\mathbf{G})$ with dimension $k$. If the channel error probability is low, the number of errors among the hard received codewords is very small; hence the rank of the hard received codeword matrix tends to be less than $n$. Therefore, we cannot find any $n$-by-$n$ full-rank matrix, so the recovery process cannot be started. A revised Canteaut-Chabaud algorithm was proposed to solve this issue by using GE to find the full-rank $n_s$-by-$M$ matrix where $n_s \leq n$. We can observe that the revised version can work for the low channel error probability scenario.  Meanwhile, it can be observed that the revised Canteaut-Chabaud algorithm and Yu's algorithm get higher recovery failure probabilities compared to our proposed GEV algorithm. This is because Yu's algorithm relies on conducting GE on an $k$-by-$n$ full-rank matrix to obtain the systematic matrix. The $k$ linear independent column in the $k$-by-$n$ full-rank matrix might not match the $k$ linear independent column in the true generator matrix $\textbf{G}$, and then we will get the wrong parity check matrix. Besides, by using only the same $N_{GE}$, the complexity of the proposed GEV algorithm is much lower than that of Yu's algorithm. Moreover, we can also find that using the proposed GEV algorithm, the recovery failure probability can be reduced significantly when a larger $N_{GE}$ is used. In particular, by using only $N_{GE}=10$, the proposed GEV algorithm with known and unknown noise variance have a lower recovery failure probability as compared with the original Canteaut-Chabaud algorithm, the 
revised Canteaut-Chabaud algorithm, and Yu's algorithm. Moreover, the proposed GEV algorithm with known noise variance has a better recovery performance than that with unknown noise variance. This is because the proposed GEV algorithm relies on the correct noise variance $\sigma^{2}$ to calculate the threshold $T_{\mathbf{h}}$ can be calculated with \eqref{eq:valueT} and the probability $p_{\mathbf{h}_{t}}$ by \eqref{probability-3}. When there is an error in the estimation of noise variance $\sigma^{2}$, some obtained correct dual words will be deemed as incorrect ones and will be removed from the correct dual words candidate list,  resulting in the code recovery failure.

\begin{figure}[!t]
	\setlength{\abovecaptionskip}{-0.05cm}
\centering \includegraphics[width=8cm]{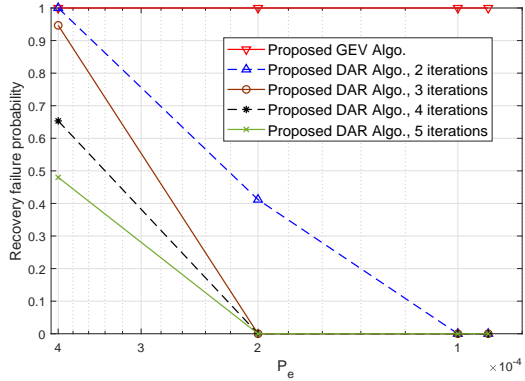} \caption{The recovery failure probabilities of the proposed GEV algorithm and
the proposed DAR algorithm for the LDPC (648, 324) code at different
values of $P_{e}$.}
\label{errorBSC-3} 
\end{figure}

Next, the performance of the proposed DAR algorithm is verified by simulations. The LDPC parity-check matrices defined in the IEEE 802.11n standard are used in our simulations \cite{11n09wifi}. Accordingly, we use codewords of length $648$. The parity-check matrix is specified corresponding to code rate $R=1/2$ in \cite{11n09wifi}. The value of $M$ is set to be 5000, and the recovery failure probabilities of the proposed GEV algorithm and the proposed DAR algorithm with different iterations are shown in Fig. \ref{errorBSC-3}. From Fig. \ref{errorBSC-3}, it can be observed that for all the $P_{e}$ tested, part of the dual words can be recovered in the first iteration, and more dual words can be recovered in the subsequent iterations. In summary, after each iteration, there are decreases in the recovery failure probability, and finally, all the dual words can be recovered. 

\begin{figure}[!t]
	\setlength{\abovecaptionskip}{-0.05cm}
\centering \includegraphics[width=8cm]{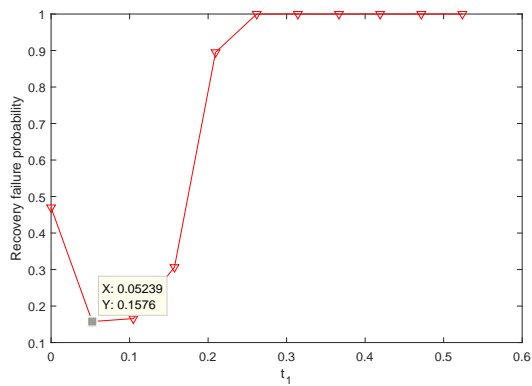} \caption{The recovery failure probabilities of the proposed EDCSDAR algorithm for the LDPC (648, 540) code with different
error detection parameters of $t_{1}$.}
\label{errorBSC-11} 
\end{figure}

\begin{figure}[!t]
	\setlength{\abovecaptionskip}{-0.05cm}
\centering \includegraphics[width=8cm]{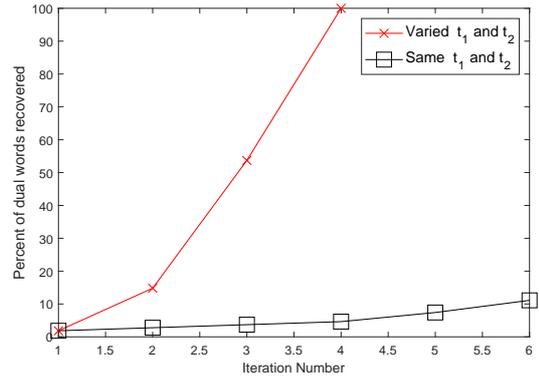} \caption{The percentage of dual words recovered of the proposed EDCSDAR algorithm with the same detection parameters as compared with the proposed EDCSDAR algorithm with varied detection parameters for the LDPC (648, 540) code at different iterations.}
\label{errorBSC-14} 
\end{figure}
Then, we carry out the simulation for the LDPC (648, 540) code. The number of received codewords $m$ is $3,840,000$. The BER is $1.5\times 10^{-3}$. From Fig. \ref{errorBSC-11}, it is clear that  there is an optimal detection parameter set. The optimal value is $0.16$ when we set $t_{1}=0.05$ and $t_{2}=0.28$, which is obtained with monte carlo simulation when $M_s$ is set to be $600$. From Fig. \ref{errorBSC-14}, we can observe that in the first iteration, when only two dual words are recovered, the new BERs are not significantly reduced after the decoding. However, when we use the same detection threshold parameters, more than $1400$ codewords do not have any detected error in them and we only need a little more than $k=540$ codewords with no detected errors to conduct the dual word recovery process with the GEV method. In this case, we should revise the error detection threshold $t_1$ for the first sub-matrix until only a little more than $k=540$ codewords with no detected errors are obtained. With this change of the detection threshold $t_1$, we can obtain the codewords with fewer true errors. In the following iterations, we use the same strategy--we revise the error detection threshold $t_1$ for the first sub-matrix until only a little more than $k=540$ codewords with no detected errors are obtained. This detection parameter adjustment will continue until all the dual words are recovered. The performance of the EDCSDAR with the same detection parameters and varied parameters are compared in Fig. \ref{errorBSC-14}. We can observe that by adjusting the detection parameter, only $4$ iterations are needed to recover the LDPC (648, 540) code.


\section{\label{sec:Conclusion}Conclusion}

This paper investigates the problem of blind recovery of the parity check matrix of linear block codes and proposes a fast recovery scheme consisting of 4 parts. Firstly, this scheme performs initial error position detection among the received codewords and selects the desirable codewords.  Then, this scheme conducts Gaussian elimination on a $k$-by-$k$ full-rank matrix. Then, it uses a threshold and the reliability associated to verify the recovered dual words. Finally, it performs decoding with partially recovered dual words. These three parts can be combined into different schemes for different noise level scenarios. The GEV that combines Gaussian elimination and verification has a significantly lower recovery failure probability and a much lower computational complexity than an existing Canteaut-Chabaud-based algorithm. The DAR and EDCSDAR schemes can improve the code recovery performance over GEV for high noise level scenarios, and their computational complexities remain much lower than the Canteaut-Chabaud-based algorithm.


\section*{Appendix A \label{sec:Appendix-A}}

\section*{Existing Algorithms}
\subsection{Brute Force Search}

In \cite{Cluzeau06B}, an algorithm for finding such dual words by using
brute force search (BFS) was proposed, in which the received
noisy bitstream is firstly divided into $M$ blocks of length $n$
and then arranged in an $M$-by-$n$ binary matrix $\mathbf{X}$.

According to (\ref{innerhc-1}), for any $\mathbf{h}$ belonging to
$\mathcal{C}^{\bot}$, when there is no transmission error, we have
$\mathbf{h}\mathbf{X}^{T}=\mathbf{0}$; when the channel is noisy,
we have $\mathbf{h}\mathbf{X}^{T}=\mathbf{w_{h}}$, where $\mathbf{w_{h}}$
is a vector of length $M$. According to the definition of $P_{e}$
and $\mathbf{r}_{v}$ in Equations (4) and (5), it can be derived that 
\begin{equation}
\text{Pr}(\mathbf{h}\mathbf{r}_{v}^{T}=0)=\frac{1+(1-2P_{e})^{\left|\mathbf{h}\right|}}{2}\label{eq:7}
\end{equation}
and 
\begin{equation}
\text{Pr}(\mathbf{h}\mathbf{r}_{v}^{T}=1)=\frac{1-(1-2P_{e})^{\left|\mathbf{h}\right|}}{2},\label{eq:8}
\end{equation}
where $\left|\mathbf{h}\right|$ denotes the Hamming weight of $\mathbf{h}$.
Thus, when $\mathbf{h}$ is a dual word in $\mathcal{C}^{\bot}$,
the Hamming weight $\left|\mathbf{w_{h}}\right|$ of $\mathbf{w_{h}}$
has a binomial distribution with a mean value of $\mu_{1}=\frac{M}{2}(1-(1-2P_{e})^{\left|\mathbf{h}\right|})$
and a variance of $\sigma_{1}^{2}=\frac{M}{4}(1-(1-2P_{e})^{2\left|\mathbf{h}\right|})$;
when $\mathbf{h}$ is not a dual word in $\mathcal{C}^{\bot}$, $\left|\mathbf{w_{h}}\right|$
has a binomial distribution with a mean value of $\mu_{2}=\nicefrac{M}{2}$
and a variance of $\sigma_{2}^{2}=\nicefrac{M}{4}$. It can be obtained that a threshold
$T_{\mathbf{h}}$ can be used to separate the two distributions and
estimate whether $\mathbf{h}$ belongs to $\mathcal{C}^{\bot}$. To
achieve a high detection accuracy, the intersection between the two
distributions should be small. This can be achieved by using $M$
larger than a certain value. To make the distance from $T_{\mathbf{h}}$
to $\mu_{1}$ and $\mu_{2}$ larger than three times the standard
deviations, the value of $M$ should satisfy 
\begin{equation}
M>\left(\frac{3\left(\sqrt{1-(1-2P_{e})^{2\left|\mathbf{h}\right|}}+1\right)}{(1-2P_{e})^{\left|\mathbf{h}\right|}}\right)^{2},\label{valueM}
\end{equation}
and the corresponding threshold $T_{\mathbf{h}}$ can be derived as 
\begin{eqnarray}
T_{\mathbf{h}} & = & \frac{M}{2}\left(1-\frac{(1-2P_{e})^{\left|\mathbf{h}\right|}}{2}\right)\nonumber \\
 &  & +3\frac{\sqrt{M}}{4}\left(\sqrt{1-(1-2P_{e})^{2\left|\mathbf{h}\right|}}-1\right).\label{eq:valueT}
\end{eqnarray}

\subsection{Code Recovery Using the Canteaut-Chabaud Algorithm}

The aforementioned BFS algorithm can hardly be used for code block length $n$ larger than $20$, shorter than most existing linear block codes. In order to deal with these codes, an algorithm based on the Canteaut-Chabaud information set decoding algorithm \cite{Canteaut98A} was proposed by Valembois \cite{Valembois01D} and recently were used by Cluzeau \cite{Cluzeau09R,Cluzeau06B} and Cote \cite{Cote09R} to recover different types of codes. This Canteaut-Chabaud algorithm includes the following steps: 
\begin{enumerate}
\item Select $n$ linearly independent rows in the $M$-by-$n$ matrix $\mathbf{X}$ to form a $n$-by-$n$ matrix $\mathbf{N}$. 
\item Perform GE on this $n$-by-$n$ full-rank matrix $\mathbf{N}$ by swapping and XORing columns of $\mathbf{X}$. After $\mathbf{N}$ becomes an identity matrix, we will get a new matrix $\mathbf{X}'$. Store the transition matrix $\mathbf{Q}$ such that $\mathbf{X}\mathbf{Q}=\mathbf{X}'$. 
\item Choose a small window of $l$ rows among the $M-n$ remaining rows of $\mathbf{X}'$. 
\item Use the same technique as in the BFS algorithm to find all combinations of $2p$ columns XORing to $\mathbf{0}$ on the $l$ rows of the window, where $p$ is usually equal to 1 or 2. 
\item For each set of $2p$ columns which meets the requirement in Step 4, verify that the XOR on the columns of $\mathbf{X}'$ is of weight lower than a threshold $T$, where $T=\frac{M}{2}-\sqrt{\frac{Mn\ln2}{2}}$. 
\item Each word $\mathbf{h}$ of weight lower than $T$ can be converted to a dual word $\mathbf{h}'$ by $\mathbf{h}'=\mathbf{h}\mathbf{Q}$. 
\end{enumerate}
When performing GE on this $n$-by-$n$ full-rank matrix for several iterations, the Canteaut-Chabaud algorithm only changes one position in the previous iteration information set to make the GE step less costly \cite{Canteaut98A}. 

\subsection{Code Recovery Using the Revised Canteaut-Chabaud Algorithm}

The aforementioned Canteaut-Chabaud algorithm cannot be used at low channel error probability. This is because the Canteaut-Chabaud algorithm relies on conducting GE on an $n$-by-$n$ full-rank matrix. For a ($n,k$) linear block code, all the codeword is among a vector space span($G$) with dimension $k$. If the channel error probability is low, the number of errors among the hard received codewords is very small. Hence the rank of the hard received codeword matrix tends to be less than $n$. For an error-free scenario, the rank of the hard received codeword is only $k$. Therefore, we cannot find any $n$-by-$n$ full-rank matrix, so the recovery process cannot be started. In order to deal with low error and error-free scenarios, a revised Canteaut-Chabaud algorithm was proposed. This revised Canteaut-Chabaud algorithm includes the following steps: 
\begin{enumerate}
\item Perform GE on the $n$-by-$M$ matrix $\mathbf{X}^T$ by swapping and XORing rows of $\mathbf{X}^T$. After $\mathbf{X}^T$ becomes an upper trapezoidal matrix, we will get a new matrix $\mathbf{X}'$. Store the transition matrix $\mathbf{Q}_1$ such that $\mathbf{Q}_1\mathbf{X}^T=\mathbf{X}'$. If the rank of matrix $\mathbf{X}^T$ is $n_s,n_s \leq n$, the last $n-n_s$ rows of matrix $\mathbf{X}^T$ are all zeros row vectors.
\item Each word $\mathbf{h}$ in the last $n-n_s$ rows of matrix $\mathbf{Q}_1$ is a dual word.
\item Select $n_s$ not-all-zero rows in the $n$-by-$M$ matrix $\mathbf{X}'$ to form a $n_s$-by-$M$ matrix $\mathbf{X}_s$.
\item Select $n_s$ linearly independent rows in the $n_s$-by-$M$ matrix $\mathbf{X}_s$ to form a $n_s$-by-$n_s$ matrix $\mathbf{N}$. 
\item Perform GE on this $n_s$-by-$n_s$ full-rank matrix $\mathbf{N}$ by swapping and XORing rows of $\mathbf{X}_s$. After $\mathbf{N}$ becomes an identity matrix, we will get a new matrix $\mathbf{X}'_s$. Store the transition matrix $\mathbf{Q}_2$ such that $\mathbf{Q}_2\mathbf{X}_s=\mathbf{X}'_s$. 
\item Choose a small window of $l$ rows among the $M-n_s$ remaining rows of $\mathbf{X}'_s$. 
\item Use the same technique as in the BFS algorithm to find all combinations of $2p$ columns XORing to $\mathbf{0}$ on the $l$ rows of the window, where $p$ is usually equal to 1 or 2. 
\item For each set of $2p$ columns which meets the requirement in Step 4, verify that the XOR on the columns of $\mathbf{X}'_s$ is of weight lower than a threshold $T$, where $T=\frac{M}{2}-\sqrt{\frac{Mn\ln2}{2}}$. 
\item Each word $\mathbf{h}$ of weight lower than $T$ can be converted to a dual word $\mathbf{h}'$ by $\mathbf{h}'=[\mathbf{h},\mathbf{0}_{1\times n_s}]\mathbf{Q}_2\mathbf{Q}_1$. 
\end{enumerate}

\subsection{Code Recovery Using the Algorithm
of Finding Low Weight code-words}

Moreover, an algorithm based on finding low weight codewords \cite{Canteaut98A} was proposed by Yu \cite{Yu17L} to recover LDPC codes. Yu's work focused on the recovery of LDPC codes which has a special property of sparse parity-check matrix. This algorithm includes the following steps: 
\begin{enumerate}
\item Select $M_c$ rows randomly in the $M$-by-$n$ matrix $\mathbf{X}$ to form a $M_c$-by-$n$ matrix $\mathbf{X}_c$. 
\item Select $k'$ linearly independent rows in the $M_c$-by-$n$ matrix $\mathbf{X}_c$
to form a $k'$-by-$n$ matrix $\mathbf{C'}$.  
\item Perform GE on this $k'$-by-$n$ matrix $\mathbf{C'}$ by
swapping and XORing rows of $\mathbf{C'}$. After swapping column of $\mathbf{C'}$ to obtain the systematic matrix $\mathbf{G}_{sys}^{'}=\left(\mathbf{I}_{k'},\mathbf{P}\right)_{I'}$, where $I'$ is the column swapping/permutation pattern to gather $k'$ linear independent column into the first $k'$ columns, then we will get a new  parity check matrix $\mathbf{H}_{sys}^{'}=\left(\mathbf{P}^T,\mathbf{I}_{n-k'}\right)_{I'}$.
\item Find $n-k'$ low weight dualwords from $\mathbf{H}_{sys}^{'}$  for LDPC codes using Canteaut-Chabaud algorithm with performing GE on this $k'$-by-$k'$ full-rank matrix for $N_{c1}$ iterations. 
\item Proceed to Step 1 until $N_{c2}$ iterations have reached.
\end{enumerate}

\subsection{Computational Complexity}

In a code recovery scenario, it is very important that the proposed scheme is feasible for a code with a large code length $n$. Further, from the above description we see that the computational complexity of the existing algorithms is mainly related to the BFS and GE processes. For the BFS algorithm, the computational complexity is $O(2^{n}\cdot M)$ which is exponential with the code length $n$. For the Canteaut-Chabaud algorithm, the computational complexity is $O(M\cdot n^{2})$ for step 2 and $O\left(\left(_{\:\:p}^{n/2}\right)^{2}\right)$ for step 4. Steps 1 to 5 will be repeated $N_{GE}$ times. Hence, the computational complexity of this Canteaut-Chabaud algorithm is $O(N_{GE}\cdot (M\cdot n^{2}+\left(_{\:\:p}^{n/2}\right)^{2}))$. From the above description, we can see that the computational complexity of the Canteaut-Chabaud algorithm is much less than that of the the BFS algorithm. However, it is still very high for the code with code length $n\geq1000$. For Yu's algorithm, the computational complexity is $O(M\cdot n^{2})$ for steps 2 and 3 and $O\left(N_{c1}\cdot\left[n\cdot (n-k)^2+\left(_{\:\:p}^{(n-k)/2}\right)^{2}\right)\right])$ for step 4. Steps 1 to 5 will be repeated $N_{c2}$ times. Hence, the total complexity is $O\left(N_{c2} \left[M n^{2}+\left(N_{c1}\left(n (n-k)^2+\left(_{\:\:p}^{(n-k)/2}\right)^{2}\right)\right)\right]\right)$.

\section*{Appendix B\label{sec:Appendix-B}}

\section*{Gaussian Elimination based Recovery Algorithms}

\subsection{Recovery of the First $n-k$ Dual Word Candidates}

Recall that the received noisy hard bitstream is firstly divided into $M$ blocks of length $n$ and then arranged row-wise into an $M$-by-$n$ binary matrix $\mathbf{X}$. We can divide the matrix $\mathbf{X}$ into two sub-matrices $\mathbf{X}_{1}$ with size $M$-by-$k$ and $\mathbf{X}_{2}$ with size $M$-by-$(n-k)$, so that $\mathbf{X}=\left(\mathbf{X}_{1},\mathbf{X}_{2}\right)$. Then we select $k$ row vectors from the matrix $\mathbf{X}$ where corresponding row vectors in the sub-matrix $\mathbf{X}_{1}$ are linearly independent to form a $k$-by-$n$ matrix $\mathbf{R}$, which is given by 
\begin{eqnarray}
\mathbf{R} & \!\!\!\!=\!\!\!\! & \left(\mathbf{r}_{1}^{T},\ldots,\mathbf{r}_{k}^{T}\right)^{T}=\left(\!\!\!\begin{array}{cccc}
r_{1}^{1} & r_{1}^{2} & \cdots & r_{1}^{n}\\
r_{2}^{1} & r_{2}^{2} & \cdots & r_{2}^{n}\\
\vdots & \ddots & \ddots & \vdots\\
r_{k-1}^{1} & r_{k-1}^{2} & \cdots & r_{k-1}^{n}\\
r_{k}^{1} & r_{k}^{2} & \cdots & r_{k}^{n}
\end{array}\!\!\!\right).\label{eq:4.1.1}
\end{eqnarray}

When there is no transmission error, for the matrix $\mathbf{R}$, we have 
\begin{equation}
\mathbf{R}=\mathbf{S}\mathbf{G},\label{eq:4.1.2}
\end{equation}
where $\mathbf{R}$ is the received hard codeword matrix, $\mathbf{S}=\left(\mathbf{s}_{1}^{T},\mathbf{s}_{2}^{T},\ldots,\mathbf{s}_{k}^{T}\right)^{T}$ is the $k$-by-$k$ message matrix and $\mathbf{G}$ is the $k$-by-$n$ generator matrix defined in Section II. The matrix $\mathbf{R}$ can be divided into two sub-matrices $\mathbf{R}_{1}$ with size $k$-by-$k$ and $\mathbf{R}_{2}$ with size $k$-by-$(n-k)$, so that $\mathbf{R}=\left(\mathbf{R}_{1},\mathbf{R}_{2}\right)$. The $j^{th}$ column in the sub-matrix $\mathbf{R}_{2}$ is defined as $\mathbf{r}_{2,j}=(r_{2,j}^{1},...,r_{2,j}^{k})^{T}$, and $\mathbf{r}_{2,j}$ can be obtained by 
\begin{eqnarray}
\mathbf{r}_{2,j} & = & \mathbf{S}\times\mathbf{p}_{j}=\mathbf{R}_{1}\times\mathbf{p}_{j},\label{eq:4.2.2}
\end{eqnarray}
where $\mathbf{p}_{j}=(p_{j}^{1},...,p_{j}^{k})^{T}$ is the $j^{th}$ column of $\mathbf{P}$.

When the $k$ message vectors among the $k$-by-$k$ message matrix $\mathbf{R}_{1}$ are linearly independent, i.e., $\mathbf{R}_{1}$ is a full-rank matrix, we have 
\begin{eqnarray}
\mathbf{\hat{p}}_{j} & = & \left(\mathbf{R}_{1}\right)^{-1}\mathbf{r}_{2,j}.\label{eq:4.1.3}
\end{eqnarray}
where $\widehat{\mathbf{p}}_{j}$ is the recovered $\mathbf{p}_{j}$.

According to (\ref{eq:4.1.3}), with the knowledge of the full-rank matrix $\mathbf{R}_{1}$, we would be able to obtain the $j^{th}$ column of the parity check matrix $\mathbf{P}$. Hence, we need to find $k$ codewords among the $M$ codewords of $\mathbf{X}$ whose message vectors are linearly independent to form $\mathbf{R}$ and obtain $\left(\mathbf{R}_{1}\right)^{-1}$ simultaneously. To achieve this, we adapt the Gauss\textendash Jordan elimination through pivoting (GJETP) algorithm \cite{Golub96M} to perform GE on the sub-matrix $\mathbf{X}_{1}$.

By adapting GJETP, the $k$ linearly independent rows among the $M$ rows of $\mathbf{X}_{1}$ can be found efficiently. Meanwhile, the inverse matrix of $\mathbf{R}_1$ can be obtained without performing matrix inversion. More specifically, our proposed dual word recovery algorithm includes the following steps:
\begin{enumerate}
\item Perform GE on the $k$-by-$M$ matrix $\mathbf{X}_{1}^{T}$ by swapping and XORing rows of $\mathbf{X}_{1}^{T}$ to obtain a new matrix $\mathbf{A}$ which is row echelon form of $\mathbf{X}_{1}^{T}$ and store the transition matrix $\mathbf{D}_{t1}$ such that $\mathbf{A}=\mathbf{D}_{t1}\mathbf{X}_{1}^{T}$. (Note that the elementary row operations do not affect the dependence relations between the column vectors. This makes it possible to use row reduction to find linearly independent column vectors.)
\item Store the column indexes of the columns with pivots in a full-rank index table ($\mathbf{FRIT}$) since the independent columns of the reduced row echelon form $\mathbf{A}$ are precisely the columns with pivots \cite{Golub96M}.
\item Form the full-rank matrix $\left(\mathbf{R}_{1}\right)^{T}$ with the columns vectors whose column indexes are in the $\mathbf{FRIT}$.
\item Do back substitution on $\mathbf{A}$ to obtain the corresponding transition matrix $\mathbf{D}_{t1}$ as the inverse matrix of the full-rank matrix $\left(\mathbf{R}_{1}\right)^{T}$. 
\end{enumerate}
We now obtain $\mathbf{D}_{t1}=\left(\left(\mathbf{R}_{1}\right)^{T}\right)^{-1}$ and $\left(\mathbf{R}_{1}\right)^{-1}=\left(\mathbf{D}_{t1}\right)^{T}$\footnote{Note that the transpose of an invertible matrix is also invertible, and its inverse is the transpose of the inverse of the original matrix, i.e., $\left(\left(\mathbf{U}\right)^{T}\right)^{-1}=\left(\left(\mathbf{U}\right)^{-1}\right)^{T}$.}. The details of this algorithm are presented in Algorithm 1.

\subsection{Recovery of Additional Dual Word Candidates}

We need to perform GE on different $k$-by-$k$ full-rank matrices to obtain more dual word candidates. GE is computationally expensive. In order to avoid repeating this procedure, we follow the idea proposed in \cite{Canteaut98A} to choose at each step to replace only one column of the previously obtained full-rank matrix and operate on this new matrix.

Recall that $\mathbf{FRIT}$ is the previously obtained index vector which contains the $k$-element subset of column indexes of $\mathbf{X}_{1}^{T}$ and the columns indexed by $\mathbf{FRIT}$ are linearly independent. $\left[\mathbf{X}_{1}^{T}\right]_{(:,\mu)}$ and $\left[\mathbf{X}_{1}^{T}\right]_{(:,\mathbf{\mathbf{FRIT}}\backslash\{\lambda\}\})}$ are linearly independent if and only if $\mathbf{A}_{(x,\mu)}=1$ where $x$ is the non-zero element row index number of $\mathbf{A}_{(:,\lambda)}$. After finding the new linearly independent column with column index $\mu$, we can obtain a new set of column indexes $\mathbf{\mathbf{FRIT}}'=\mathbf{\mathbf{FRIT}}\backslash\{\lambda\}\cup\{\mu\}$ to form the full-rank matrix $\mathbf{R}_{1}'$. We also need to obtain the inverse matrix of the newly obtained full-rank matrix $\mathbf{R}_{1}'$. This can be done by simply adding the $x$-th row of matrix $\mathbf{A}$ to all other rows $\mathbf{A}_{(z,:)}$ when the corresponding element $\mathbf{A}_{(z,\mu)}$ is not zero.

Hence, after $k$ steps, we can obtain a completely new full-rank matrix and its inverse matrix. Then, we can calculate $\mathbf{\hat{p}}_{j},1\leq j\leq n-k$, with (\ref{eq:4.1.3}).

\subsection{Verification of the Recovered Dual Word Candidates}

When the channel is noisy, the recovered dual words $\mathbf{h}_{1},\mathbf{h}_{2}...$, might not be correct. These recovered dual words are rows of the obtained parity check matrix $\hat{\mathbf{H}}=\left(\hat{\mathbf{P}},\mathbf{I}_{(n-k)}\right)$, where $\hat{\mathbf{P}}=\left(\hat{\mathbf{p}}_{1},\cdots,\hat{\mathbf{p}}_{n-k}\right)$ is the recovered $\mathbf{P}$ matrix. To improve the accuracy of the recovery, we propose to perform verification on all the recovered dual words candidates based on the reliability of each recovered dual word.

To measure the reliability of each dual word, the probability $p_{\mathbf{h}_{t}}$, which denotes the probability that a tested $n$-tuple is a true dual word. In \cite{Valembois01D}, a way to calculate the probability $p_{\mathbf{h}_{t}}$ of a tested $n$-tuple being a true dual word is proposed. When $\mathbf{h}_{t}\mathbf{X}^{T}=\mathbf{w}_{\mathbf{h}_{t}}$ and $\left|\mathbf{w}_{\mathbf{h}_{t}}\right|=d_{\mathbf{h}_{t}}$, $p_{\mathbf{h}_{t}}$ is given by 
\begin{eqnarray}
p_{\mathbf{h}_{t}} & = & \Pr\left[\left|\mathbf{w}_{\mathbf{h}_{t}}\right|=d_{\mathbf{h}_{t}}\right]\nonumber \\
 & = & \left[\text{Pr}(\mathbf{h}_{t}\mathbf{r}_{v}^{T}=0)\right]^{M-d_{\mathbf{h}_{t}}}\left[\text{Pr}(\mathbf{h}_{t}\mathbf{r}_{v}^{T}=1)\right]^{d_{\mathbf{h}_{t}}}.\label{eq:probability-2}
\end{eqnarray}

For channel with error probability $P_{e}$, inserting  (\ref{eq:7}) and (\ref{eq:8}) into (\ref{eq:probability-2}), we obtain that 
\begin{equation}
p_{\mathbf{h}_{t}}=\frac{\left(1+(1-2P_{e})^{\left|\mathbf{h}_{t}\right|}\right)^{M-d_{\mathbf{h}_{t}}}\left(1-(1-2P_{e})^{\left|\mathbf{h}_{t}\right|}\right)^{d_{\mathbf{h}_{t}}}}{2^{M}}.\label{probability-3}
\end{equation}

In summary, our proposed dual word verification scheme includes the following steps: 
\begin{enumerate}
\item For each obtained dual word \textbf{$\mathbf{h}_{t}$}, calculate $\mathbf{w}_{\mathbf{h}_{t}}=\mathbf{h}_{t}\mathbf{X}^{T}$, where $\mathbf{w}_{\mathbf{h}_{t}}$ has length $M$ and weight $d_{\mathbf{h}_{t}}$. 
\item If $d_{\mathbf{h}_{t}}\leq T_{\mathbf{h}}$, where $T_{\mathbf{h}}$ can be calculated with equation (\ref{eq:valueT}), we calculate the probability $p_{\mathbf{h}_{t}}$ of $\mathbf{h}_{t}$ by (\ref{probability-3}).
\item Use $\mathbf{h}_{t}$ and $p_{\mathbf{h}_{t}}$ to update the dual word table ($\mathbf{DWT}$) which stores maximally ($n-k$) candidates of dual words. These dual word candidates must be linearly independent. 
\item After all obtained dual word candidates are tested, take the $n-k$ dual word candidates in $\mathbf{DWT}$ as the recovered dual words and use them to form the parity check matrix. 
\end{enumerate}
To make the recovery failure probability as small as possible, the dual word candidates should be stored in table in order of their probabilities, i.e., $p_{\mathbf{h}_{i_{1}}}\geq p_{\mathbf{h}_{i_{2}}}\geq\ldots p_{\mathbf{h}_{i_{n-k}}}$.

After the verification, to make sure that the newly obtained dual word candidates are not correlated with the previously obtained dual word candidates in table $\mathbf{DWT}$, we need to use GE again. For this purpose, we propose another efficient GE algorithm, which allows us to use the knowledge of previously obtained linearly independent dual word candidates through pivoting. The details of this algorithm are presented in Algorithm 2.

\subsection{Efficient Gaussian Elimination Algorithms}

In this paper, we need to efficiently find $k$ codewords among the $M$ codewords of $\mathbf{X}$ whose message vectors are linearly independent to form $\mathbf{R}$ and obtain $\left(\mathbf{R}_{1}\right)^{-1}$ simultaneously. To achieve this, we adapt the well-known Gauss\textendash Jordan elimination through pivoting (GJETP) algorithm \cite{Golub96M} to perform GE on the sub-matrix $\mathbf{X}_{1}$. The details of this algorithm are presented in Algorithm 1.

\begin{algorithm}[!t]
\textbf{Input}: $\mathbf{X}_{1}$,

\textbf{Output}:$\mathbf{A}$, $\mathbf{D}_{t1}$, $\mathbf{\mathbf{FRIT}}$, 
\begin{enumerate}
\item Initialization:$\mathbf{A}\leftarrow\left(\mathbf{X}_{1}\right)^{T}$;
$\mathbf{D}_{t1}\leftarrow\mathbf{I}_{n}$; 
\item \textbf{for} $j$ from $1$ to $k$ do

\begin{enumerate}
\item \textbf{for} $z$ from $j$ to $M$ do

\begin{enumerate}
\item \textbf{if} the column vector $\left[\mathbf{A}\right]_{(j:k,z)}$
is not all 0 \textbf{then}

\begin{enumerate}
\item Permute the $z^{th}$ row with $(d+j-1)^{th}$ row of $\mathbf{A}$
and $\mathbf{D}_{t1}$ where $d$ is the index of the first non-zero
element of $\left[\mathbf{A}\right]_{(j:k,z)}$; 
\item $\mathbf{\mathbf{FRIT}}_{(j)}=z$; 
\end{enumerate}
\item \textbf{end if} 
\end{enumerate}
\item \textbf{end} \textbf{for} 
\item \textbf{for} $i$ from index of the non-zero entry of $\mathbf{A}_{((d+1):k,z)}$

\begin{enumerate}
\item Apply the XOR operation to the $z^{th}$ row and $i^{th}$ row of
$\mathbf{A}$ and $\mathbf{D}_{t1}$ in order to have $\mathbf{A}_{(i,z)}$
= 0: 
\end{enumerate}
\item \textbf{end} \textbf{for} 
\end{enumerate}
\item \textbf{end for} 
\item \textbf{for} $j$ from $k$ to $2$ do

\begin{enumerate}
\item \textbf{for} $i$ from index of the non-zero entry of $\mathbf{A}_{(1:(j-1),\mathbf{\mathbf{FRIT}}_{(j)})}$

\begin{enumerate}
\item Apply the XOR operation to the $i^{th}$ row and $j^{th}$ row of
$\mathbf{A}$ and $\mathbf{D}_{t1}$ to have $\mathbf{A}_{(i,\mathbf{\mathbf{FRIT}}_{(j)})}$
= 0: 
\end{enumerate}
\item \textbf{end} \textbf{for} 
\end{enumerate}
\item \textbf{end for} 
\end{enumerate}
\textbf{return} $\mathbf{A}$, $\mathbf{D}_{t1}$, $\mathbf{\mathbf{FRIT}}$,

\caption{\label{alg:Efficient-Gaussian-Elimination-2}Efficient Gaussian Elimination
Algorithm to obtain the Full-Rank Matrix and Its Inverse Matrix}
\end{algorithm}

For the obtained dual word candidates, to verify the linear independence between the newly obtained dual word candidates and the previously obtained dual word candidates, we propose another algorithm that allows us to use the knowledge of previously obtained linearly independent dual word candidates. The details of this algorithm are presented in Algorithm 2.

At the beginning, a dual word candidate $\mathbf{h}_{1}$ is obtained and verified with the threshold in step 2). When $\mathbf{h}_{1}\neq\mathbf{0}$, we let $\mathbf{Q}=(\mathbf{h}_{1}^{T})$ and conduct GE on the $k$-by-$1$ matrix $\mathbf{Q}$ by swapping and adding rows to obtain a new matrix $\mathbf{Q}'=[1,0,...,0]^{T}$, which is the row echelon form of $\mathbf{Q}$, and the corresponding transition matrix is $\mathbf{D}_{t2}$. Now, we already have one dual word candidate. Then when another dual word candidate $\mathbf{h}_{2}^{T}$ is obtained, to make sure that $\mathbf{h}_{2}^{T}$ is linearly independent with $\mathbf{h}_{1}^{T}$, we can calculate that $\mathbf{h}_{2}'=\mathbf{D}_{t2}\mathbf{h}_{2}^{T}$ and check whether there is a new pivot appeared in $\mathbf{h}_{2}'$ when compared with $\mathbf{Q}'$ \cite{Golub96M}. If there is, we conclude that $\mathbf{h}_{2}^{T}$ is linearly independent to $\mathbf{h}_{1}^{T}$ and combine $\mathbf{h}_{1}'$ and $\mathbf{h}_{2}'$ to form an $n$-by-$2$ matrix. Then we conduct GE on the $n$-by-$2$ matrix $\mathbf{Q}=\left(\mathbf{Q}',\mathbf{h}_{2}'\right)$ by swapping and adding rows to obtain the row echelon form of $\mathbf{Q}$, i.e., $\mathbf{Q}'$, and the corresponding updated transition matrix $\mathbf{D}_{t2}$. After processing a number of obtained dual word candidates, suppose that we already have $N_{i}$ dual words in the $\mathbf{DWT}$, and an $n$-by-$N_{i}$ full-rank matrix $\mathbf{Q}'$ which has row echelon form and the corresponding updated transition matrix $\mathbf{D}_{t2}$. When a new dual word candidates $\mathbf{h}_{t}$ is obtained, we calculate $\mathbf{h}_{t}'=\mathbf{D}_{t2}\mathbf{h}_{t}^{T}$ and check whether a new pivot \cite{Golub96M} appears in $\mathbf{h}_{t}'$ when compared with $\mathbf{Q}'$. If there is, we conclude that $\mathbf{h}_{t}$ is linearly independent to the $N_{i}$ obtained linearly independent dual word candidates in the $\mathbf{DWT}$, we then calculate the new row echelon form matrix with size $n$ by $(N_{i}+1)$ and the corresponding updated transition matrix $\mathbf{D}_{t2}$. If there is no new pivot, we keep $\mathbf{Q}'$ and $\mathbf{P}'$ to be unchanged. 
\begin{algorithm}[!t]
\textbf{Input}: $\mathbf{x}$, $N_{R}$, $\mathbf{Q}$, $\mathbf{D}_{t2}$

\textbf{Output}: $\mathbf{Q}'$, $\mathbf{D}_{t2}$', $N_{R}$ 
\begin{enumerate}
\item Initialization: $\mathbf{q}'=\mbox{mod}(\mathbf{P}\cdot\mathbf{\mathbf{q}},2)$;
$j=$$N_{R}$$+1$; 
\item \textbf{if} the column vector $\mathbf{q}'_{(j:n)}$ is not all 0
\textbf{then}

\begin{enumerate}
\item Permute the $j^{th}$ row with $(d+j-1)^{th}$ row of $\mathbf{q}$
and $\mathbf{D}_{t2}$ where $d$ is the index of the first non-zero
element of $\mathbf{q}'_{(j:n)}$; $N_{R}=N_{R}+1$; $\mathbf{Q}'=(\mathbf{Q},\mathbf{q}')$; 
\item \textbf{for} $i$ is among index of the non-zero element of $\mathbf{q}'_{((d+j):m)}$

\begin{enumerate}
\item Apply the XOR operation to the $(i+d+j-1)^{th}$ row and $j^{th}$
row $\mathbf{Q}'$ and $\mathbf{D}_{t2}$ in order to have $\mathbf{Q}_{((i+d+j-1),j)}^{'}$
= 0: 
\end{enumerate}
\item \textbf{end} \textbf{for} 
\end{enumerate}
\item \textbf{end if} 
\item $\mathbf{D}_{t2}'=\mathbf{D}_{t2}$ 
\end{enumerate}
\textbf{return} $\mathbf{Q}'$, $\mathbf{D}_{t2}'$, $N_{R}$

\caption{\label{alg:Efficient-Gaussian-Elimination-1-1}Gaussian Elimination
Algorithm to test the Linear Independence of the Column Vectors}
\end{algorithm}

\end{document}